\documentclass{article}
\usepackage{graphics} 
\usepackage{epsfig}

\begin{document}

\begin{center}{\Large\bf The Modified Nonlinear Schr\"odinger Equation:\\ Facts and
Artefacts}\\
\medskip
E.V. Doktorov\\
\medskip
B.I. Stepanov Institute of Physics, 220072 Minsk, Belarus\\
\medskip

Abstract\\
\end{center}
We argue that the integrable modified nonlinear Schr\"odinger
equation with the nonlinearity dispersion term is the true
starting point to analytically describe subpicosecond pulse
dynamics in monomode fibers. Contrary to the known assertions,
solitons of this equation are free of self-steepening and the
breather formation is possible.\\
\medskip

\noindent
{\bf 1. Introduction}\\
\medskip

Soliton-based optical communication systems serve as an exciting
example of the application of a purely mathematical concept
(soliton) to modern technology. The nonlinear Schr\"odinger
equation (NSE)
\begin{equation}
iu_t+(1/2)u_{xx}+|u|^2u=0
\end{equation}
is the adequate model to describe picosecond soliton evolution in
monomode fibers \cite{Has-Tap}. Here $u(x,t)$ is the envelope of
the pulse electric field and coordinates $t$ and $x$ measure
distance along the fiber and time in a frame comoving with the
pulse group velocity, respectively. The applicability of NSE
depends crucially on the assumption that the spatial width of the
envelope is much larger that the carrier wavelength. Besides, the
success of this model is substantially related to integrability of
NSE \cite{Z-S} and hence to the controllability of soliton
parameters \cite{Has}. Various more subtle effects accompanying
the picosecond soliton propagation are usually treated as a
perturbation of the integrable model.

On the other hand, dynamics of subpicosecond optical pulses
($\le100$ fs) is not well governed by NSE because  the above
mentioned assumption is not satisfied. The spectral width of
subpicosecond pulses becomes comparable with the carrier
frequency, and three main additional effects - nonlinearity
dispersion, intrapulse Raman stimulated scattering and linear
higher-order dispersion - should be taken into account \cite{Agr}:
\begin{equation}
iu_t+(1/2)u_{xx}+|u|^2u=i\alpha_1\left(|u|^2u\right)_x+\alpha_2\left(|u|^2
\right)_xu+i\alpha_3u_{xxx}.
\end{equation}
The terms in rhs of (2) account for the above additional effects.
In general,  extra terms violate integrability of the equation.
Hence, a question can be posed: does there exist an equation that
will be integrable as NSE and at the same time would be more
relevant in the subpicosecond range? The answer is positive
because the modified NSE (MNSE)
\begin{equation}
iu_t+(1/2)u_{xx}+|u|^2u+i\alpha\left(|u|^2u\right)_x=0, \quad
\alpha\in\rm{Re}
\end{equation}
with the $\alpha$-dependent nonlinearity dispersion term is still
integrable though the associated spectral problem is more
involved than the Zakharov-Shabat one. Namely, the initial-value
problem for MNSE (3) can be solved within the framework of the
Wadati-Konno-Ichikawa (WKI) spectral problem \cite{WKI}. A
careful study of the WKI spectral problem (or the quadratic
bundle) for MNSE and related equations was undertaken by
Gerdjikov and Ivanov \cite{GI}. Explicit soliton solutions to
MNSE obtained in \cite{Vys} and \cite{Rao} turned out too
complicated for practical use. That is the reason that MNSE is
usually treated as NSE with $\alpha$-dependent perturbation term,
especially as actual values of $\alpha$ are normally small.

Among the other things, such treatment gave rise to some
misunderstanding. First, it was shown in \cite{Lis} that the
initially symmetric hyperbolic-secant pulse evolving in
accordance with MNSE (3) develops an asymmetric self-phase
modulation and a self-steepening. There is a wide-spread opinion
that the self-steepening  is an inherent property of the
subpicosecond pulse dynamics \cite{Agr} that should be minimized
for proper operation of an information system. Second, it is well
known \cite{Sats} that the initial pulse $2\,{\rm{sech}}\,x$
evolving according to NSE produces the NSE breather (the bound
state of two solitons). On the other hand, the same initial pulse
decays into separate solitons when evolving according to MNSE.
Hence, it was inferred that MNSE does not admit breathers (or
higher-order solitons) \cite{Gol}.

Our aim here is to show that the situation with subpicosecond
soliton dynamics is rather different. We argue that the
integrable MNSE is the true starting point to analytically
describe this dynamics. It is remarkable that numerical
simulations of the MNSE-based soliton propagation revealed
various regimes which cannot be explained by treating the
$\alpha$-dependent term in (3) as a perturbation of NSE
\cite{Oh}. We derive the MNSE soliton solution that is
non-perturbative w.r.t. $\alpha$ and demonstrate that the MNSE
soliton propagates without any self-steepening. Besides, we
explicitly obtain breather solution to MNSE. Numerical
simulations confirm the stability of the MNSE breather.\\
\medskip

\noindent {\bf 2. MNSE soliton}\\
\medskip

We will employ the Riemann-Hilbert (RH) problem \cite{Nov} for
solving nonlinear equations. Let us start with the Lax pair for
MNSE (3):
\begin{equation}
\Psi_x = \Lambda(k)\left[ \sigma_3,\Psi \right] + 2ikQ\Psi, \quad
\end{equation}
\begin{equation}
\Psi_t=\Omega(k)\left[\sigma_3,\Psi\right]
+\biggl(\frac{4i}{\alpha}k^3Q+2ik^2Q^2\sigma_3 \nonumber
-\frac{i}{\alpha} kQ+kQ_x\sigma_3-2i\alpha kQ^3\biggr)\Psi,
\nonumber
\end{equation}
\[
\Lambda(k)=-\frac{2i}{\alpha} \left( k^2-\frac{1}{4} \right),\quad
\Omega(k)=-\frac{4i}{\alpha^2}\left(k^2-\frac{1}{4}\right)^2.
\nonumber
\]
Here the Hermitian matrix $Q=\pmatrix{0 &u\cr \bar u& 0\cr}$
stands for the potential of the spectral problem (4), $k$ is a
spectral parameter. The standard procedure is:\\
a) building the Jost solutions of the linear spectral problem
(4);\\
b) building the solutions $\Phi_\pm$ which are analytical in
complementary regions of the $k$-plane;\\
c) formulation of the RH problem for $\Phi_\pm$ with the standard
normalization
\begin{equation}
\Phi_\pm\rightarrow I \qquad {\rm for} \quad |k|\to\infty,
\end{equation}
where $I$ is the identity matrix.

It is, however, easy to see by substituting the asymptotic
expansion w.r.t. to $k^{-1}$ of $\Psi$ to the spectral problem (4)
that this problem does not agree with the standard normalization.
On the other hand, an associated equation with the fifth-order
nonlinearity,
\begin{equation}
iv_t+(1/2)v_{xx}-i\alpha v^2\bar v_x+|v|^2v+\alpha^2|v|^4v=0
\end{equation}
has the Lax pair as well with the WKI spectral problem
\begin{eqnarray}
\Psi_x^{(A)}&=&\Lambda(k)[\sigma_3,\Psi^{(A)}]+(2ikQ_A+i\alpha
Q_A^2\sigma_3)\Psi^{(A)}, \\ Q_A&=&\pmatrix{0 &v\cr \bar v& 0\cr}
\nonumber\end{eqnarray} that agrees with the standard
normalization, and with the same dispersion relation $\Omega(k)$
for the temporal Lax equation. Moreover, equations (3) and (6)
are gauge equivalent and solutions of MNSE (3) follow from those
of (6) by means of a simple algebraic relation
\begin{equation}
Q=g^{-1}Q_Ag, \qquad g(x,t)=\Psi^{(A)}(k=0,x,t).
\end{equation}
The associated equation (6) does not have such an obvious
physical interpretation as the MNSE but it has an extremely
simple soliton solution. Hence, we will not solve  MNSE directly.
Instead we will integrate the associated equation (6) and then
will obtain solutions of MNSE by the algebraic relation (8).

We begin with the spectral problem (7) for the associated
equation. At first we define the Jost solutions $J_\pm$,
$J_\pm\rightarrow I$ at $x\to \pm\infty$, which are interrelated
with the scattering matrix $S$, $J_-E=J_+ES$. Here
$E=\exp\left(\Lambda(k)x\right)$. Dividing the Jost solutions
into columns, $J_\pm=(J^{(1)}_\pm,J^{(2)}_\pm)$, it can be shown
by the standard analysis of integral equations that the columns
$J^{(1)}_+$ and $J^{(2)}_-$ are analytical in the first and third
quadrants of the $k$-plane. Hence, the matrix function
$\Phi_+=(J^{(1)}_+,J^{(2)}_-)$ is analytical as a whole in the
same quadrants. The matrix $\Phi_+$ can be expressed in terms of
the Jost solution $J_+$ and some entries of the scattering matrix:
\[
\Phi_+=J_+ES_+E^{-1}, \qquad S_+=\pmatrix{1&s_{12}\cr 0& s_{22}}.
\]
Because the potential $Q_A$ is Hermitian, we have the involutions
$ \left[J_\pm(k)\right]^\dagger=\left[J_\pm(\bar k)\right]^{-1}$,
$S(k)^\dagger=S(k)^{-1}$.  They allow us to introduce the matrix
function $\Phi_-$, $\Phi_-^{-1}(k)=\Phi_+(\bar
k)^\dagger=ES_+^\dagger E^{-1}J_+^{-1}$ that is analytical in the
second and forth quadrants. Thereby, we can pose the RH problem
with the standard normalization,
\begin{equation}
\Phi_-^{-1}\Phi_+=EG(k)E^{-1}, \qquad \Phi_\pm\rightarrow I \quad
{\rm at} \quad k\to\infty,
\end{equation}
where
\[
G(k)=S_+^\dagger S_+=\pmatrix{1&s_{12}\cr {\bar s}_{12}&1}, \quad
k\!\in\!\!\left\{k=\xi-i\eta, \,\,\, \xi\eta=0\right\},
\]
as a problem of analytical factorization of the non-degenera\-te
matrix $G(k)$ defined on both the real and imaginary axes of the
$k$-plane.

In general, the function $\Phi_+$ has zeros in some points $k_j$
lying in the first and third quadrants, $\det\Phi_+(k_j)=0$.
Hence, in these points there exist eigenvectors $|n_j\rangle$ with
zero eigenvalue. It is important that zeros $k_j$ appear by pairs
$(k_j,-k_j)$. It is a feature of the WKI spectral problem. Hence,
the single soliton of the associated equation  is determined by
two zeros $k_1$ and $-k_1$. The zeros $k_j$, eigenvectors
$|n_j\rangle$ and the matrix $G(k)$ comprise the RH data. Because
we deal with the solitons only, $G(k)$ being related with the
continuous spectrum of the spectral problem, is taken to be the
identity matrix.

If $\Phi_+$ is a solution of the RH problem (9), it can be
expanded in the asymptotic series
$\Phi_+=I+\Phi^{(1)}_+/k+\Phi^{(2)}_+/k^2+\cdots$. Substituting
this expansion into the spectral problem (7), we  reconstruct the
potential $Q_A$:
\begin{equation}
Q_A=(1/\alpha)[\sigma_3,\,\Phi^{(1)}_+]=(2/\alpha)\sigma_3\Phi^{(1)}_+.
\end{equation}

Now we derive a soliton of the associated equation (6). Let us
have zeros $k_1$ and $-k_1$ and two eigenvectors $|n_\pm\rangle$.
It can be easily shown that the eigenvectors obey the equations
\[
|n_+\rangle_x=\Lambda(k_1)\sigma_3|n_+\rangle, \qquad
|n_+\rangle_t=\Omega(k_1)\sigma_3|n_+\rangle.
\]
Hence, we obtain explicit space and time dependencies of the
eigenvectors,
\begin{eqnarray}
|n_+\rangle&=&\left(\begin{array}{c}\exp[\Lambda(k_1)x+\Omega(k_1)t]\exp(a+i\phi_0)\\
\exp[-\Lambda(k_1)x-\Omega(k_1)t]\end{array}\right), \nonumber \\
|n_-\rangle&=&\sigma_3|n_+\rangle, \quad \langle
n_\pm|=|n_\pm\rangle^\dagger.\nonumber
\end{eqnarray}
Here $a,\phi_0={\rm const}$. It can be shown by the dressing
method \cite{Nov} that the matrix $\Phi_+$ is represented as
($k_\pm\equiv\pm k_1$)
\begin{equation}
\Phi_+(k)=I-\sum_{j,l=\pm}\frac{|n_j\rangle(D^{-1})_{j\,l}\langle
n_l|}{k-\bar k_l}, \qquad D_{j\,l}=\frac{\langle
n_j\,|n_l\rangle}{k_l-\bar k_j},
\end{equation}
\[
\Phi_-^{-1}(k)=I+\sum_{j,l=\pm}\frac{|n_j\rangle(D^{-1})_{j\,l}\langle
n_l|}{k-k_j}.
\]
Because the eigenvectors are known explicitly, we can evaluate
the matrix $\Phi_+$ as $\Phi_+(k)=I-D_+/(k-\bar k_1)-D_-/(k+\bar
k_1)$, where
\begin{eqnarray}
D_+&=&\frac{k_1^2-\bar k_1^2}{2}
\left(\begin{array}{cc}e^z(k_1e^{-z}+{\bar
k}_1e^z)^{-1} & e^{i\varphi}(k_1e^{-z}+{\bar k}_1e^z)^{-1}\\
e^{-i\varphi}(k_1e^z+{\bar k}_1e^{-z})^{-1} & e^{-z}(k_1e^z+{\bar
k}_1e^{-z})^{-1}\end{array}\right), \nonumber \\
D_-&=&-\sigma_3D_+\sigma_3. \nonumber
\end{eqnarray}
We introduced here new independent variables $z$ and $\varphi$:
\[
z=-(1/w)(x-Vt-x_0), \quad \varphi=Vx-(1/2)(V^2-w^{-2})t+\varphi_0,
\]
\[
x_0=aw, \quad \varphi_0={\rm const},
\]
where the soliton velocity $V$ and width $w$ (see below) are
defined by
\[
V=\frac{1}{2\alpha}\left(1-2(k_1^2+\bar k_1^2)\right),\qquad
w=\frac{1}{2i}\frac{\alpha}{k_1^2-\bar k_1^2}.
\]
Hence, the eigenvalue $k_1$ is expressed in terms of velocity and
width as
\begin{equation}
k_1=(1/2)(1-\alpha V-i\alpha/w)^{1/2}, \qquad {\rm Im}k_1<0.
\end{equation}
Expanding then $\Phi_+$ in the asymptotic series, we obtain from
Eq. (10) the soliton solution of the associated equation:
\begin{equation}
v_s=\frac{i}{w}\frac{e^{i\varphi}}{k_1e^{-z}+{\bar k}_1e^z}.
\end{equation}
It has indeed a very simple form.

An important aspect of the solution (13) should be noted. Namely,
the parameter $\alpha$ that enters the soliton width $w$ appears
in the denominator. Hence, we cannot reproduce the soliton (13)
considering the associated equation (6) as the $\alpha$-perturbed
NSE. Nevertheless, there exists a procedure \cite{My} to perform
the limit $\alpha\to 0$. Namely, representing $k_1$ as
$k_1=(1/2)-(\alpha/2)\lambda_1+{\cal O}(\alpha^2)$, we obtain in
this limit from Eq. (13) the NSE soliton with the eigenvalue
$\lambda_1$.

As regards the MNSE soliton $u_s$, it follows from $v_s$ by means
of the algebraic relation (8). Indeed, $u_s=(g_2/g_1)v_s$,
\[
\left(\begin{array}{cc}g_1 & 0\\
0 & g_2 \end{array}\right)=\Phi_+(k=0)=I+\frac{2}{\bar
k_1}\left(\begin{array}{cc}D_{+11} & 0\\ 0 & D_{+22}
\end{array}\right).
\]
Explicitly we have
\[
g_1=\frac{k_1}{\bar k_1}\frac{k_1e^z+\bar
k_1e^{-z}}{k_1e^{-z}+\bar k_1e^z}, \quad g_2=\frac{k_1}{\bar
k_1}\frac{k_1e^{-z}+\bar k_1e^z}{k_1e^z+\bar k_1e^{-z}},
\]
\begin{equation}
u_s=\frac{i}{w}\frac{k_1e^{-z}+{\bar k}_1e^z}{(k_1e^z+{\bar
k}_1e^{-z})^2}e^{i\varphi}.
\end{equation}
The MNSE soliton (14) looks much simpler than those derived in
\cite{Vys} and \cite{Rao}. In the limit $\alpha\to 0$, the
solitons of both the MNSE and the associated equation reproduce
one and the same NSE soliton.

Square of module of $u_s$ (14) is written as
\[
|u_s|^2=\frac{1}{w^2}(k_1e^{-z}+{\bar k}_1e^z)^{-1}(k_1e^z+{\bar
k}_1e^{-z})^{-1}
\]
\[
=\frac{1}{2w^2}\left[1-\alpha V+\sqrt{(1-\alpha
V)^2+\frac{\alpha^2}{w^2}}\,{\rm
cosh}\frac{2}{w}(x-Vt)\right]^{-1}.
\]
We see from this relation that the envelope $|u_s|$ moves holding
its shape, i.e., without any self-steepening.\\
\medskip

\noindent
{\bf 3. MNSE breather}
\medskip

To derive the MNSE breather, we start from four zeros $\pm k_1$
and $\pm k_2$, where $k_j=(1/2)(1-\alpha V_j-i\alpha/w_j)^{1/2},
\quad{\rm Im}k_j <0 $ (cf. Eq. (12)). Because we seek for a bound
state of two solitons, we put $V_1=V_2\equiv V$ and, without loss
of generality, $V=0$. We have four eigenvectors with the property
$|n_{2j}\rangle=\sigma_3|n_{2j-1}\rangle$, $j=1,2$. Namely,
\begin{eqnarray}
|n_1\rangle&=&\left(\begin{array}{c}\exp[-(x/2w_1)+i(t/4w_1^2)+i\varphi_{11}]
\\
\exp[(x/2w_1)-i(t/4w_1^2)+i\varphi_{12}]\end{array}\right),\nonumber\\
|n_3\rangle&=&\left(\begin{array}{c}\exp[-(x/2w_2)+i(t/4w_2^2)+i\varphi_{21}]
\\ \exp[(x/2w_2)-i(t/4w_2^2)+i\varphi_{22}]\end{array}\right)
\end{eqnarray}
with a special relation for the constant phases
$\varphi_{11}-\varphi_{12}-\varphi_{21}+\varphi_{22}=\pi$. Denote
$\lambda_{2j-1}\equiv k_j$ and $\lambda_{2j}\equiv-k_j$, $j=1,2$.
Then the matrix function $\Phi_+(k)$ for the breather is written
as (cf. Eq. (11))
\[
\Phi_+(k)=I-\sum_{m,n=1}^4\frac{|n_m\rangle(D^{-1})_{m\,n}\langle
n_n|}{k-\bar\lambda_n}, \quad D_{m\,n}=\frac{\langle
n_m\,|n_n\rangle}{\lambda_n-\bar\lambda_m}.
\]
We omit cumbersome but evident calculations performed along the
lines of Sect. 2 and give below the explicit expression for the
MNSE breather at rest:
\begin{eqnarray}
u_{br}&=&(g_2'/g_1')v_{br}, \\
v_{br}&=&\frac{w_1-w_2}{w_1+w_2}D^{-1}\Bigl[w_1\left(k_1e^{x/w_1}+\bar
k_1e^{-x/w_1}\right)e^{it/2w_2^2} \nonumber \\
&+&w_2\left(k_2e^{x/w_2}+\bar
k_2e^{-x/w_2}\right)e^{it/2w_2^2}\Bigr], \nonumber\\
D&=&w_1w_2\left(k_1e^{x/w_1}+\bar k_1e^{-x/w_1}\right)
\left(k_2e^{x/w_2}+\bar k_2e^{-x/w_2}\right)
\nonumber \\
&-&w_+^2\left(k_1e^{x/w_+-it/w_+w_-}-\bar
k_2e^{-x/w_++it/w_+w_-}\right) \nonumber \\
&\times&\left(k_2e^{x/w_++it/w_+w_-}-\bar
k_1e^{-x/w_+-it/w_+w_-}\right), \nonumber \\
g_1'&=&\Phi_{+11}(k=0)=1-\frac{i\alpha}{2D}
\Biggl[\frac{w_1-w_2}{w_1+w_2}\left( w_1\frac{\bar k_1}{\bar
k_2}-w_2\frac{\bar k_2}{\bar
k_1}\right)e^{-2x/w_+}\nonumber \\
&+&w_1\frac{k_1}{\bar k_2}e^{2x/w_-}+w_2\frac{k_2}{\bar
k_1}e^{-2x/w_-} \nonumber \\
&+&w_+\left(\frac{k_1}{\bar k_1}e^{-2it/w_+w_-}+\frac{k_2}{\bar
k_2}e^{2it/w_+w_-}\right)\Biggr], \nonumber \\
g_2'&=&\Phi_{+22}(k=0)=1-\frac{i\alpha}{2\bar D}
\Biggl[\frac{w_1-w_2}{w_1+w_2}\left( w_1\frac{\bar k_1}{\bar
k_2}-w_2\frac{\bar k_2}{\bar k_1}\right)e^{2x/w_+}\nonumber \\
&+&w_1\frac{k_1}{\bar k_2}e^{-2x/w_-}+w_2\frac{k_2}{\bar
k_1}e^{2x/w_-} \nonumber \\
&+&w_+\left(\frac{k_1}{\bar k_1}e^{2it/w_+w_-}+\frac{k_2}{\bar
k_2}e^{-2it/w_+w_-}\right)\Biggr]. \nonumber \\
\nonumber\end{eqnarray} Here $w_\pm^{-1}=(1/2)\left(w_1^{-1}\pm
w_2^{-1}\right)$. It is seen that the MNSE breather oscillates
with the period $T=\pi w_+w_-$ and reproduces in the limit
$\alpha\to0$ the well known NSE breather \cite{Sats}. Fig. 1 shows
the square of module of the breather solution (16) for $w_1=1/3$,
$w_2=1$ and $\alpha=0.1$. We see that there is no any decay of
the MNSE breather.\\
\medskip

\noindent
{\bf Conclusion}
\medskip

We consider MNSE as a natural integrable generalization of  NSE
to the range of subpicosecond optical pulses. It is shown in this
paper that MNSE possesses  the basic ingredients (solitons and
breathers) of integrable nonlinear equations. To justify the
applicability of these results to the description of actual
subpicosecond pulses, we should account for at least the
intrapulse Raman scattering that breaks integrability of the
equation. A possibility to reduce an adverse action of this
effect is discussed in \cite{Afa} on the basis of the
perturbation theory for the MNSE soliton \cite{My}. A novel way
to suppress the Gordon-Haus effect for the MNSE soliton was
revealed in \cite{Kut}. Recently quasiradiation solution of a
compound model including MNSE was obtained by Zabolotskii
\cite{Zab}.
\begin{figure}

  \includegraphics{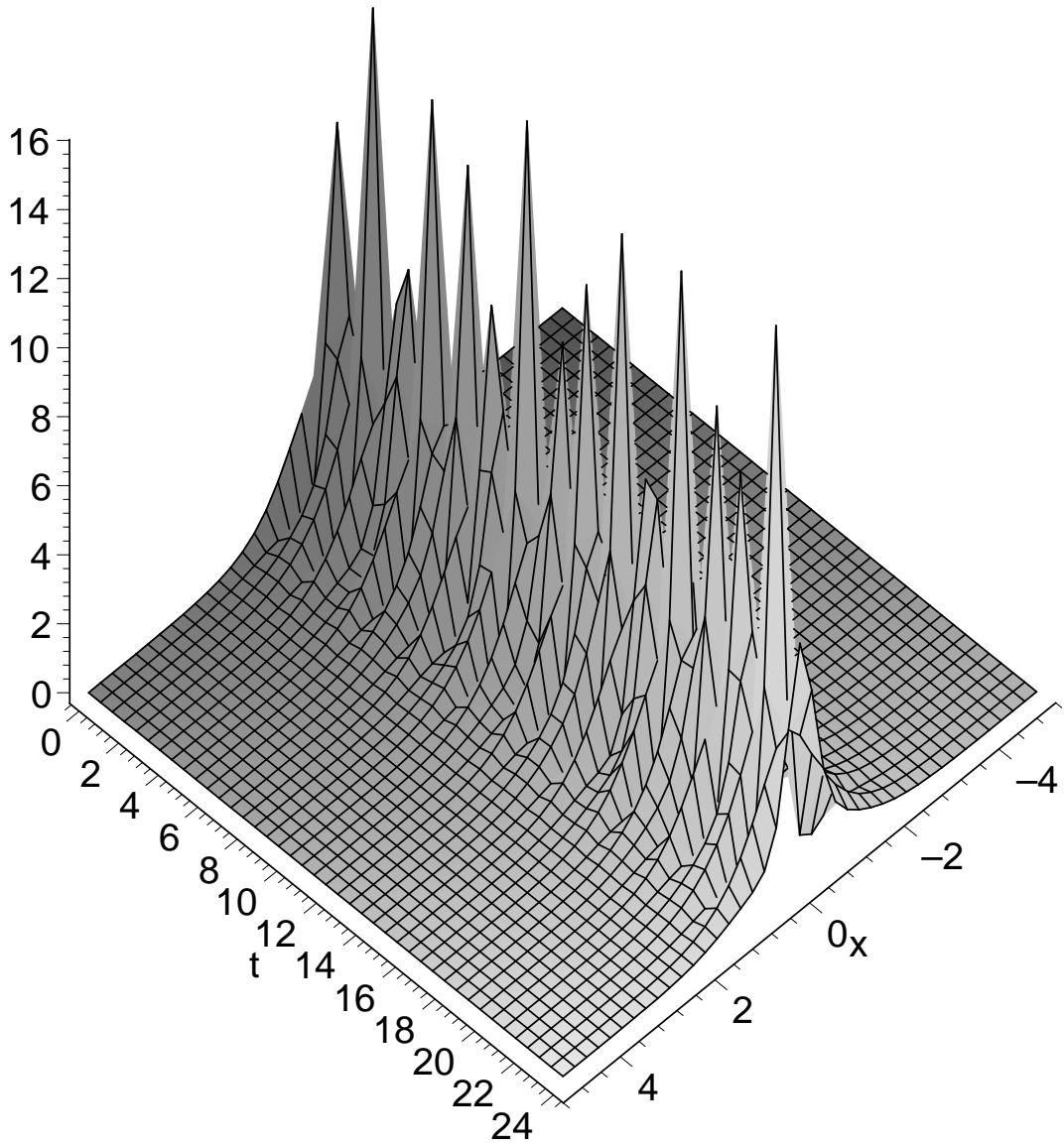}

\caption{Square of module of the MNSE breather solution (16) for
$w_1=1/3$, $w_2=1$ and $\alpha=0.1$.} \label{fig:1}
\end{figure}
\medskip

\noindent
{\bf Acknowledgements}
\medskip

The author is grateful to the Organizing Committee of the
conference GIN'01 (Bansko, Bulgaria) for financial support.
Stimulating discussions with Prof. V.S. Gerdjikov and Dr. V.S.
Shchesnovich are greatly appreciated. The author thanks V.S.
Shchesnovich for invaluable help with the illustrative material.
This work was partly supported by grant no. 97-2018 from
INTAS-Belarus.

\end{document}